\documentclass[a4paper,fleqn]{cas-sc}

\usepackage[numbers]{natbib}
\usepackage{float}
\usepackage{subfig}
\def\tsc#1{\csdef{#1}{\textsc{\lowercase{#1}}\xspace}}
\tsc{WGM}
\tsc{QE}
\tsc{EP}
\tsc{PMS}
\tsc{BEC}
\tsc{DE}

\begin{document}
\let\WriteBookmarks\relax
\def\floatpagepagefraction{1}
\def\textpagefraction{.001}

\shorttitle{The acoustoelastic simulation}

\shortauthors{Xu Maoyu et~al.}

\title [mode = title]{The weak form method for acoustoelastic simulation with arbitrary prestress}                      

\author[1]{Xu Maoyu}
                        

\credit{Conceptualization, Methodology, Software, Validation, Formal analysis, Investigation, Writing – original draft, Data curation, Investigation}

\affiliation[1]{organization={College of Sciences},
    addressline={Northeastern University}, 
    city={Shenyang},
    postcode={110819}, 
    country={China}}


\author[2]{Changsheng Liu}
\credit{Resources, Supervision, Funding acquisition}

\affiliation[2]{organization={Key Laboratory for Anisotropy and Texture of Materials Ministry of Education},
	addressline={Northeastern University}, 
	city={Shenyang},
	postcode={110819}, 
	country={China}}

\author[1]{Yu Zhan}[orcid=0000-0002-1723-2612]
\credit{Visualization, Resources, Writing – review \& editing, Funding acquisition}
\cormark[1]

\ead{zhanyu@mail.neu.edu.cn}

\cortext[cor1]{Corresponding author}

\begin{abstract}
Acoustoelastic theory has been widely used to evaluate the residual stress (or prestress). However, most of the research remains focused on plate under simple tensile stress condition.  In this paper, we propose a new approach: using weak form PDE modeling for acoustoelastic simulation. In the theory part, the weak form of acoustoelastic theory and semi-analytical finite element (SAFE) method is derived. In the numerical simulation part, two cases: the propagation direction is perpendicular to the prestress and parallel to the prestress are presented. The results are compared with the superposition of partial bulk wave (SPBW) method and the previously commonly used Effective Elastic Constants (EEC) method. The results show that the method proposed in this paper is highly accurate. Compared to the EEC method, this approach has no theoretical flaws and aligns more closely with the theoretical solution, especially in the low frequency range where the EEC method's results are almost incorrect. This study provides an accurate method for the simulation of acoustoelasticity.
\end{abstract}


\maketitle

\section{Introduction}
Acoustoelastic theory is the theory that studies the relationship between wave propagation and stress states. Hughes and Kelly \cite{PhysRev.92.1145} experimentally determine the relationship between the velocities of longitudinal and shear waves and applied stress, and are the first to propose using the third-order elastic constants \cite{Sneddon1954FiniteDO} to study acoustoelastic theory. The general equations for a small displacement superimposed on a finite deformation of a perfectly elastic material of arbitrary symmetry are derived anew by Toupin and Bernstein \cite{10.1121/1.1908623}. Pao \cite{Pao1985AcoustoelasticWI} derived the theory of acoustoelastic in detail and systematically, studied the propagation of ultrasound in orthotropic elastic solids with initial stress, and derived the relationship between strain and wave velocity when the coordinate system coincides with the principal strain axis. Acoustoelastic stress detection technology has been widely applied due to its advantages of non-destructive, low cost, and high accuracy. Chaki \cite{Chaki2009StressLM} used the acoustoelastic effect to measure the stress levels in prestressed steel strands. Hirao \cite{10.1063/1.1144328} proposed a practical method of acoustoelastic stress measurement based on electromagnetic acoustic resonance. Lillamand \cite{LILLAMAND2010655} studied the acoustoelastic effect in concrete material under uniaxial compressive loading. Acoustoelastic theory also has been widely used in the fields of thermal stress monitoring and Rayleigh wave stress measurement \cite{ZENG2023106948, CHEN2023106905,LIU2022106639}.

Due to the complexity of acoustoelastic theory, its current applications primarily focus on stress evaluation based on the relationship between wave velocity and the magnitude of stress. For more complex stress states, there have been fewer studies, and most of them are simple uniform tensile stress states. Due to the advantages of guided wave, such as long-distance propagation, strong adaptability, and cost-effectiveness, it is widely used in non-destructive testing and structural health monitoring. Dispersion curves of guided waves are useful for analyzing wave speeds and identifying guided wave modes. Some work has already been done on the influence of stress on the dispersion curves of guided waves. Lematre \cite{10.1121/1.2336989} established a model for guided wave propagation under prestress gradients by modifying the recursive stiffness matrix method, and analyzed the impact of the gradient on the guided wave characteristics of piezoelectric plates and films. Gandhi \cite{10.1121/1.4740491} studied the anisotropy of Lamb wave dispersion curves in isotropic media under a biaxial stress field. Pei \cite{10.1121/1.4967756} found that the change in Lamb wave velocity, due to the acoustoelastic effect, for the S1 mode exhibits about ten times more sensitive, in terms of velocity change, than the traditional bulk wave measurements, and those performed using the fundamental Lamb modes. Peddeti \cite{peddeti2018dispersion} used the semi-analytical finite element (SAFE) method to analyze the dispersion curves of Lamb wave propagation in prestressed plates. Zhengyan \cite{yang2019acoustoelastic} studied acoustoelastic guided wave propagation  in axial stressed arbitrary cross-section using the Effective Elastic Constants (EEC) method. Dispersion curves for complex cross-sections is obtained, but the stress state is limited to simple axial stress. Peng \cite{zuo2020acoustoelastic} extended the prestress state to include shear stress. Kubrusly's work on acoustoelastic theory is relatively detailed, comparing the EEC method with theoretical solutions and highlighting the shortcomings of the EEC method \cite{kubrusly2016time}. There has been less research on the propagation of waves in prestressed media. Kubrusly used the EEC method to simulate the propagation of Lamb waves. Other studies mostly involve modifying the element stiffness matrix or lack clarity regarding the internal algorithms of commercial software \cite{Pei2011NumericalSO, Ferrari2009TheAE, Loveday12}.

Most of the above research focuses on simple stress states such as tension or shear, and the method of modifying the element stiffness matrix is inherently flawed in theory. This flaw could be critical when dealing with the small effect of acoustoelasticity. However, the state of prestress is very complex in industrial applications and other situations. To solve this problem, a new method need to be proposed. 

In this paper, a method for acoustoelastic simulation under complex prestress conditions is proposed. In the theory section, the governing equation of acoustoelastic theory and its weak form is thoroughly derived. The weak form of the semi-analytical finite element method has also been derived in a similar manner. The acoustoelastic simulation is used to simulate the wave propagation process in the prestressed medium, and the SAFE is used to obtain the dispersion curve of the lamb wave. Next, the detailed variables and steps for using the COMSOL Multiphysics Weak Form PDE interface are introduced. In the validation and application of simulation part, two cases: the propagation direction is perpendicular to the prestress and parallel to the prestress are presented. In the case when the propagation direction is perpendicular to the prestress, the phase velocity difference curve caused by prestress is compared with the results from the superposition of partial bulk wave (SPBW) method, and the two results matched well. In the case when the propagation direction is parallel to the prestress, the variation curve of phase velocity with stress is compared with the EEC method. Several concerns regarding the EEC method are pointed out, including the convergence value at zero frequency, the positive and negative gains of prestress at low and high frequencies, as well as the lack of zero crossing region. Finally, the propagation of A0 mode Lamb wave in a plate is simulated, providing a reference for subsequent research. 

\section{Theory}
\subsection{Acoustoelastic theory}
For the theory of acoustoelasticity, three states of the medium are adopted: the configuration without stress and strain is called the nature state and is represented by the coordinates $\xi_{\alpha}(\alpha=1,2,3)$; the configuration under static deformation is called the initial state and is represented by the coordinates $X_{J}(J=1,2,3)$; the configuration of initial state superimposed with small acoustic perturbations is called the final state and is represented by the coordinates $x_{j}(j=1,2,3)$. Pao derived the equations of acoustoelasticity in natural frame of reference \cite{Pao1985AcoustoelasticWI}. For predeformed medium in nature configuration: 
\begin{equation}\label{eq:govering}
	\frac{\partial}{\partial\xi_\beta}P^n_{\alpha\beta}=\rho^0\frac{\partial^2u_\alpha}{\partial t^2}
\end{equation}
where
\begin{equation}
	P^n_{\alpha\beta} = A^{n}_{\alpha\beta\gamma\delta}\frac{\partial u_\gamma}{\partial\xi_\delta}
\end{equation}
\begin{equation}
	A^{n}_{\alpha\beta\gamma\delta}=s_{\beta\delta}^i\delta_{\alpha\gamma}+\Gamma_{\alpha\beta\gamma\delta}=C_{\beta\delta\zeta \eta}\varepsilon^{i}_{\zeta \eta }\delta_{\alpha\gamma}+\Gamma_{\alpha\beta\gamma\delta}
\end{equation}
\begin{equation}
	\Gamma_{\alpha\beta\gamma\delta}=C_{\alpha\beta\gamma\delta}+C_{\alpha\beta\lambda\delta}\frac{\partial u_{\gamma}^{i}}{\partial\xi_{\lambda}}+C_{\lambda\beta\gamma\delta}\frac{\partial u_{\alpha}^{i}}{\partial\xi_{\lambda}}+C_{\alpha\beta\lambda\delta\zeta\eta} \varepsilon^{i}_{\zeta\eta}
\end{equation}

The above formula is valid under the condition of small predeformation.
\begin{equation}\label{eq:small_dis1}
s_{\beta\delta}^i = C_{\beta\delta\zeta \eta}\varepsilon^{i}_{\zeta \eta }
\end{equation}
\begin{equation}\label{eq:small_dis2}
\varepsilon^{i}_{\alpha\beta}=\frac12\left(\frac{\partial u_\alpha^i}{\partial\xi_\beta}+\frac{\partial u_\beta^i}{\partial\xi_\alpha}\right)
\end{equation}
where $C_{\alpha\beta\gamma\delta}$ and $C_{\alpha\beta\lambda\delta\zeta\eta}$ are the second and third order elastic constants; $u^i$ is the predeformed displacement from the natural state to the initial state; $\varepsilon^{i}$ represents the initial Cauchy strain tensor; $s^i$ is the second Piola-Kirchhoff (P–K) prestress tensor in the nature configuration; $u$ is the incremental displacement from the initial state to the finial state; $\rho^0$ is the density in nature configuration \cite{zuo2020acoustoelastic}. 

The relationship between stress increment and strain increment is expressed as
\begin{equation}
	T_{\alpha\beta}=B_{\alpha\beta\gamma\delta}^n\frac{\partial u_\gamma}{\partial\xi_\delta}
\end{equation}
where
\begin{equation}
	B^{n}_{\alpha\beta\gamma\delta}=C_{\alpha\beta\gamma\delta}+C_{\alpha\beta\lambda\delta}\frac{\partial u_{\gamma}^{i}}{\partial\xi_{\lambda}}+C_{\alpha\beta\lambda\delta\zeta\eta} \varepsilon^{i}_{\zeta\eta}
\end{equation}

The stress boundary condition can be expressed as
\begin{equation}
	t_{\alpha}=T_{\alpha \beta} n_{\beta}=B_{\alpha \beta \gamma \delta} \frac{\partial u_{\gamma}}{\partial \xi_{\delta}} n_{\beta}
\end{equation}

The governing equation and stress boundary condition are combined using weak form, the domain $\Omega$ with boundary $\Gamma$.
\begin{equation}
	\begin{aligned}
	&\int_\Omega \delta u_{\alpha} \left[\frac{\partial}{\partial\xi_\beta}\left(A^{n}_{\alpha\beta\gamma\delta}\frac{\partial u_\gamma}{\partial\xi_\delta}\right)\right] d\Omega- \int_\Omega \delta u_{\alpha} \rho^0\frac{\partial^2u_\alpha}{\partial t^2}  d\Omega \\
	& + \int_{\Gamma} \delta u_{\alpha} t_{\alpha} d \Gamma-\int_{\Gamma} \delta u_{\alpha} n_{\beta} B^{n}_{\alpha \beta \gamma \delta} \frac{\partial u_{\gamma}}{\partial \xi_{\delta}} d \Gamma =0
	\end{aligned}
\end{equation}
The divergence theorem is applied and the following formula is obtained \cite{peddeti2018dispersion}

\begin{equation}\label{eq:weak_form}
\begin{aligned}
	&- \int_{\Omega} \frac{\partial \delta u_{\alpha}}{\partial \xi_{\beta}} A^{n}_{\alpha \beta \gamma \delta} \frac{\partial u_{\gamma}}{\partial \xi_{\delta}} d \Omega - \int_{\Omega} \delta u_{\alpha} \rho^{0} \frac{\partial^{2} u_{\alpha}}{\partial t^{2}} d \Omega  \\
	&+\int_{\Gamma} \delta u_{\alpha} t_{\alpha} d \Gamma + \int_{\Gamma} \delta u_{\alpha} n_{\beta} E^{n}_{\alpha \beta \gamma \delta} \frac{\partial u_{\gamma}}{\partial x_{\delta}} d \Gamma= 0
	\end{aligned}
\end{equation}
where 
\begin{equation}
	E^{n}_{\alpha \beta \gamma \delta} = A^{n}_{\alpha\beta\gamma\delta} - B^{n}_{\alpha\beta\gamma\delta}
\end{equation}
For simplicity, a new tensor $M^n_\alpha$ is introduced.
\begin{equation}
	M^n_\alpha = n_{\beta} E^{n}_{\alpha \beta \gamma \delta} \frac{\partial u_{\gamma}}{\partial x_{\delta}}
\end{equation}

\subsection{Acoustoelastic SAFE theory}
Due to the long transmission distance and low energy loss of guided wave, it is widely used in non-destructive testing \cite{mitra2016guided}. The Semi-Analytical Finite Element Method (SAFEM) is a numerical analysis technique that combines analytical method with the Finite Element Method (FEM). By applying analytical treatment to certain dimension while discretizing others using finite elements, SAFEM simplifies calculations and enhances efficiency, making it particularly suitable for reducing one or two dimensions in three-dimensional problems \cite{ahmad2013semi, marzani2008semi}. Assuming that the Lamb wave propagates harmonically in the $\xi_3$ direction, which is perpendicular to the cross-section, the following 3D solution can be sought in form of the amplitude expansion: \cite{yang2019acoustoelastic, comsol}
\begin{equation}
	\mathbf{u}(\xi_1,\xi_2,\xi_3,t)=\begin{bmatrix}u_{1}(\xi_1,\xi_2,\xi_3,t)\\u_{2}(\xi_1,\xi_2,\xi_3,t)\\u_{3}(\xi_1,\xi_2,\xi_3,t)\end{bmatrix}=\begin{bmatrix}U_{1}(\xi_1,\xi_2)\\U_{2}(\xi_1,\xi_2)\\U_{3}(\xi_1,\xi_2)\end{bmatrix}e^{i(\omega t - k \xi_3)}
\end{equation}
where $k$, $\omega$ represent the wavenumber, angular frequency.

So the displacement gradient is following
\begin{equation} \label{eq:gradient}
	\left\{\begin{aligned}
		&\frac{\partial u_{1}}{\partial \xi_1} = \frac{\partial u_{1}}{\partial \xi_1}, \\
		&\frac{\partial u_{1}}{\partial \xi_2} = \frac{\partial u_{1}}{\partial \xi_2}, \\
		&\frac{\partial u_{1}}{\partial \xi_3} = -ik u_{1}, \\
		&\frac{\partial u_{2}}{\partial \xi_1} = \frac{\partial u_{2}}{\partial \xi_1}, \\
		&\frac{\partial u_{2}}{\partial \xi_2} = \frac{\partial u_{2}}{\partial \xi_2}, \\
		&\frac{\partial u_{2}}{\partial \xi_3} = -ik u_{2}, \\
		&\frac{\partial u_{3}}{\partial \xi_1} = \frac{\partial u_{3}}{\partial \xi_1}, \\
		&\frac{\partial u_{3}}{\partial \xi_2} = \frac{\partial u_{3}}{\partial \xi_2}, \\
		&\frac{\partial u_{3}}{\partial \xi_3} = -ik u_{3}.
	\end{aligned}\right.
\end{equation}

Substituting Eq.~(\ref{eq:gradient}) into Eq.~(\ref{eq:govering}) and Eq.~(\ref{eq:weak_form}), the govering equation and weak form become:
\begin{equation}
\frac{\partial}{\partial\xi_\beta}P^n_{\alpha\beta}=-\rho^0\omega^2u_\alpha
\end{equation}
\begin{equation}\label{eq:SAFE_weak}
\begin{aligned}
	&- \int_{\Omega} econj\left(\frac{\partial \delta u_{\alpha}}{\partial \xi_{\beta}}\right)  P^n_{\alpha\beta} d \Omega + \int_{\Omega} \delta u_{\alpha} \rho^{0}\omega^2u_\alpha d \Omega  \\
	&+\int_{\Gamma} \delta u_{\alpha} t_{\alpha} d \Gamma + \int_{\Gamma} \delta u_{\alpha} n_{\beta} E^{n}_{\alpha \beta \gamma \delta} \frac{\partial u_{\gamma}}{\partial x_{\delta}} d \Gamma= 0
\end{aligned}
\end{equation}
where $econj\left(\frac{\partial \delta u_{\alpha}}{\partial \xi_{\beta}}\right)$ denotes the conjugate of $\frac{\partial \delta u_{\alpha}}{\partial \xi_{\beta}}$.

\section{Numerical simulation}
The simulation of  wave propagation in prestressed media using the interface of the COMSOL Multiphysics Weak Form PDE will be introduced. The specific modeling details and parameter specifications will be presented. This section is divided into two parts: acoustoelastic simulation is used to simulate the wave propagation process in the prestressed medium, and the SAFE is used to obtain the dispersion curve of the lamb wave.

 \subsection{Acoustoelastic simulation}
 Typically, the form of $A^{n}_{\alpha\beta\gamma\delta}$ is quite complex and does not satisfy the symmetry $A^{n}_{\alpha\beta\gamma\delta} = A^{n}_{\alpha\beta\delta\gamma}$. Assuming the predeformation displacement gradient is known, its components are written as
 \begin{equation}
 u^i=\begin{pmatrix}\mathrm{gradiU11}&\mathrm{gradiU12}&\mathrm{gradiU13}\\\\\mathrm{gradiU21}&\mathrm{gradiU22}&\mathrm{gradiU23}\\\\\mathrm{gradiU31}&\mathrm{gradiU32}&\mathrm{gradiU33}\end{pmatrix}
\end{equation}
 
Therefore, according to the COMSOL Weak Form PDE interface, the Weak Expressions of Eq. (\ref{eq:weak_form}) should be written as: 
\begin{center}
-P11*test(ux)-P12*test(uy)-P13*test(uz)\\
-P21*test(vx)-P22*test(vy)-P23*test(vz) \\
-P31*test(wx)-P32*test(wy)-P33*test(wz) \\
\end{center}
and
\begin{center}
rho*(-utt*test(u)-vtt*test(v)-wtt*test(w))
\end{center}
The Weak Contribution of Eq. (\ref{eq:weak_form}) should be written as: 
\begin{center}
 test(u)*M1 + test(v)*M2 + test(w)*M3
\end{center}
where rho represents the density $\rho^0$, u, v, w represent the components of $u_\alpha$, ux, uy, uz, vx, vy, vz, wx, wy, wz represent the components of deformation displacement gradient $\frac{\partial u_\alpha}{\partial\xi_\beta}$, utt, vtt, wtt represent the components of acceleration $\frac{\partial^2u_\alpha}{\partial t^2}$, test() denotes the corresponding COMSOL test function interface, P11, P12, P13, P21, P22, P23, P31, P32, P33 are the components of $P^n_{\alpha\beta}$, M1, M2, M3 are the components of $M^n_{\alpha}$, the components of $P^n_{\alpha\beta}$ and $M^n_{\alpha}$ are shown in the Appendix section.

 \subsection{Acoustoelastic SAFE simulation}
The simulation steps of the acoustoelastic SAFE are generally similar to the simulation of acoustoelastic. Some new parameters are created. The theoretical variables corresponding to the COMSOL interface are shown in Table \ref{table:SAFE}. 

 \begin{table*}[!ht]
 	\centering
 	\caption{COMSOL parameters and corresponding theoretical parameters of Acoustoelastic SAFE simulation}
 	\begin{tabular}{cccc}
 		\hline
 		COMSOL Parameters & Theoretical Parameters &Expression& Description \\ \hline
 		f0& - & 5000[Hz] & Operating frequency \\ 
 		fmax & - & 10[MHz] & Maximum frequency \\ 
 		lambda& -ik& 1[rad/m] & Complex wavenumber  \\ 
 		omega & $\omega$ & 2*pi*f0& Angular frequency\\ 
 		iomega &  i$\omega$& i*omega& Complex angular frequency \\ 		
 		ikz & ik & -lambda & Phase shifted out of plane wave number\\ 
 		kz & k & ikz/i& Out of plane wave number \\
 		kn & k& kz & Out of plane wave number \\
 		uz & $\frac{\partial u_1}{\partial\xi_3}$ & -ikz*u [1]& Displacement gradient tensor, 13-component \\
 		vz &  $\frac{\partial u_2}{\partial\xi_3}$& -ikz*v [1]& Displacement gradient tensor, 23-component\\
 		wz &  $\frac{\partial u_3}{\partial\xi_3}$& -ikz*w [1]& Displacement gradient tensor, 33-component \\\hline
 	\end{tabular}
 	\label{table:SAFE}
 \end{table*}
 
 Therefore, according to the COMSOL Weak Form PDE interface, the Weak Expressions of Eq. (\ref{eq:SAFE_weak}) should be written as: 
 \begin{center}
 -P11*test(ux)-P12*test(uy)-P13*test(-uz)\\
 -P21*test(vx)-P22*test(vy)-P23*test(-vz) \\
-P31*test(wx)-P32*test(wy)-P33*test(-wz) \\
 \end{center}
 and
 \begin{center}
-rho*iomega**2*(u*test(u)+v*test(v)+w*test(w))
 \end{center}
 The Weak Contribution of Eq. (\ref{eq:SAFE_weak}) should be written as: 
 \begin{center}
 	test(u)*M1 + test(v)*M2 + test(w)*M3
 \end{center}
 where rho, u, v, w etc. have the same meaning as in the acoustoelastic simulation.
 
 \section{ Validation and application of simulation}
This section will demonstrate the validation and application of acoustoelastic simulations through specific examples. In order to verify the correctness of theoretical derivation and finite element simulation, some simulation cases are presented and compared with other papers. 

\subsection{Wave propagation direction perpendicular to the prestress}
The simulation of wave propagation perpendicular to the prestress is first established. The Lamb wave propagates in the $\xi_3$ direction of a plate subjected to a tensile stress of magnitude Siel11 in the $\xi_1$ direction. The prestress tensor can be written as
\begin{equation}
s^i=\begin{pmatrix}\mathrm{Siel11}&0&0\\\\0&0&0\\\\0&0&0\end{pmatrix}
\end{equation}

Initial Cauchy strain tensor and predeformed displacement can be calculated using Eq. (\ref{eq:small_dis1}) and Eq.~(\ref{eq:small_dis2}). As mentioned earlier, the form of $A^{n}_{\alpha\beta\gamma\delta}$ is quite complex, However, in the case of uniaxial tension, the expression for $A^{n}_{\alpha\beta\gamma\delta}$ can be simplified \cite{kubrusly2016time}. When an aluminum plate is subjected to a tensile stress of 120 MPa in the $\xi_3$ direction, the material constants of aluminum plate is shown in Table \ref{table:AI_120}, the components of $A^{n}_{\alpha\beta\gamma\delta}$ are shown in Eq. (\ref{eq:120_A}), it is noted that $A^{n}_{\alpha\beta\gamma\delta} \ne A^{n}_{\alpha\beta\delta\gamma}$. The components of $A^{n}_{\alpha\beta\gamma\delta}$ are compared with those in \cite{kubrusly2016time}, and the results show agreement.

\begin{figure}[h]
	\centering 
	\includegraphics{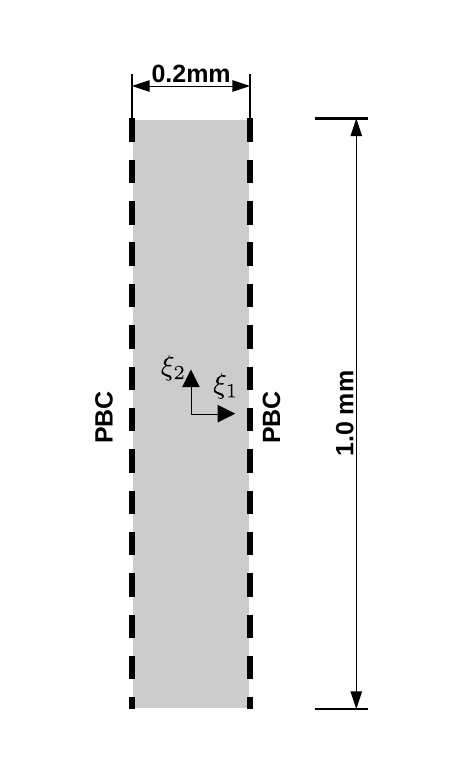} 
	\caption{The cross-sectional schematic of the aluminum plate} 
	\label{fig:AI_plate}
\end{figure}

The influence of prestress or prestrain on Lamb wave dispersion is of particular interest to us. The dispersion curve of an isotropic Aluminum 6061-T6 plate under 100 MPa tensile stress in the $\xi_1$ direction is solved using the new acoustoelastic SAFE simulation method. The material constants of aluminum plate is shown in Table \ref{table:AI_6061}. The propagation direction of the Lamb wave is along the $\xi_3$ direction. The cross-sectional schematic of the plate is shown in Fig. \ref{fig:AI_plate}. The thickness of the aluminum plate is 1.0 mm. The plate width of 0.2 mm with periodic boundary conditions (PBC) is used to replace the infinitely wide plate \cite{zuo2020acoustoelastic}.

The COMSOL "Mode Analysis" and "Parametric Sweep" study are used to calculate the Lamb wave dispersion curve. The desired number of modes is 80.  The maximum sweep frequency is 10 MHz. A total of 200 sweep analyses are performed. The same analysis is performed for the unstressed state for comparison.

\begin{figure}[h]
	\centering 
	\includegraphics{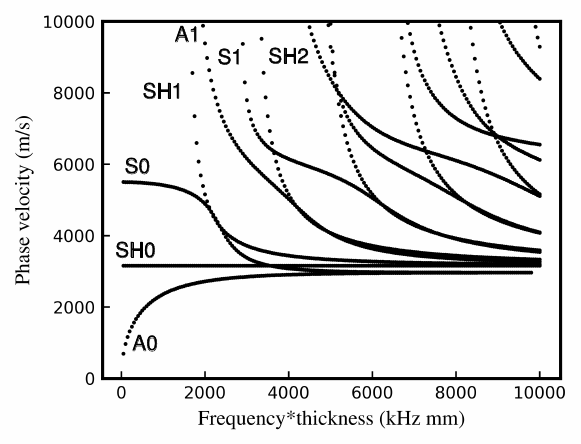} 
	\caption{The phase velocity curve of Lamb waves in the aluminum plate, with the propagation direction perpendicular to the tensile prestress 100 MPa} 
	\label{fig:disperdion}
\end{figure}

\begin{figure}[h]
	\centering 
	\includegraphics{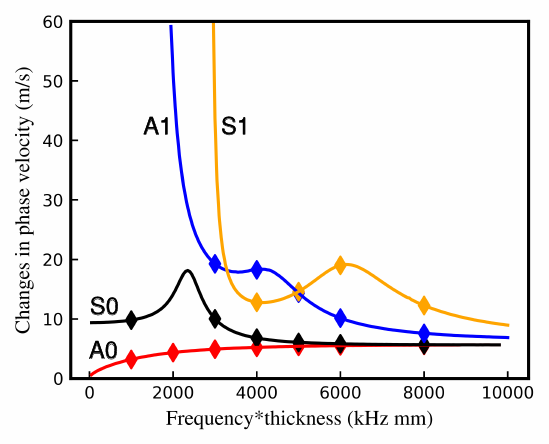} 
	\caption{The difference in phase velocity between the low order Lamb waves (A0, S0, A1, S1) in the 100 MPa tensile prestressed and unstressed states, with the propagation direction perpendicular to the tensile prestress, the diamond markers represent the results from the SPBW method} 
	\label{fig:Changes_in_phase_velocity}
\end{figure}

The phase velocity curve of the isotropic Aluminum plate under 100 MPa tensile stress in the $\xi_1$ direction is shown in Fig. \ref{fig:disperdion}. Low order Lamb waves (A0, SH0, S0, SH1, A1, S1, SH2) are labeled in Fig. \ref{fig:disperdion}. The difference in phase velocity between the low order Lamb waves (A0, S0, A1, S1) in the 100 MPa tensile prestressed and unstressed states is shown in Fig. \ref{fig:Changes_in_phase_velocity}. In Fig. \ref{fig:Changes_in_phase_velocity} , the solid line represents the phase velocity difference obtained using the acoustoelastic SAFE  method proposed in this paper, while the diamond markers represent the results from the SPBW method \cite{zuo2020acoustoelastic, 10.1121/1.4740491, 10.1121/1.4967756}. The agreement between the two methods demonstrates the accuracy and reliability of the method presented in this paper. Fig. \ref{fig:Changes_in_phase_velocity} also indicates that the acoustoelastic effect caused by stress is very small \cite{peddeti2018dispersion, Pao1985AcoustoelasticWI}.

\begin{table}[h]
	\centering
	\caption{The density, Lamé constants and Murnaghan constants of Aluminum \cite{Stobbe2005AcoustoelasticityI7}}
	\begin{tabular}{ccccccc}
		\hline
		Parameter &~& ~&Value &~&~& Units \\ \hline
		$\rho^0$ &~& ~&2700 &~&~& $\mathrm{kg/m^3}$ \\ 
		$\lambda$ &~&~& 54.9&~&~& $\mathrm{GPa}$ \\ 
		$\mu$ &~&~& 26.5 &~&~& $\mathrm{GPa}$ \\ 
		$l$ &~& ~&-252.2&~&~& $\mathrm{GPa}$ \\ 
		$m$ &~&~& -324.9&~&~& $\mathrm{GPa}$ \\ 
		$n$ &~&~& -351.2 &~&~& $\mathrm{GPa}$ \\ \hline
	\end{tabular}
	\label{table:AI_120}
\end{table}

\begin{figure*}[!ht]
	\centering
\begin{equation} \label{eq:120_A}
	A^{n} = 
	\bordermatrix{
		&11&22&33&23&13&12&32&31&21 \cr
		11&105.907&54.513&54.513&0&0&0&0&0&0\cr
		22&54.513&108.241&55.065&0&0&0&0&0&0\cr
		33&54.513&55.065&108.241&0&0&0&0&0&0\cr
		23&0&0&0&26.588&0&0&26.588&0&0\cr
		13&0&0&0&0&26.31&0&0&26.25&0\cr
		12&0&0&0&0&0&26.31&0&0&26.25\cr
		32&0&0&0&26.588&0&0&26.588&0&0\cr
		31&0&0&0&0&26.25&0&0&26.31&0\cr
		21&0&0&0&0&0&26.25&0&0&26.31\cr
	} GPa
\end{equation}
\end{figure*}

\begin{table}[!h]
	\centering
	\caption{The density, Lamé constants and Murnaghan constants of Aluminum 6061-T6 \cite{10.1063/1.1709077}}
	\begin{tabular}{ccccccc}
		\hline
		Parameter &~& ~&Value &~&~& Units \\ \hline
		$\rho^0$ &~& ~&2704 &~&~& $\mathrm{kg/m^3}$ \\ 
		$\lambda$ &~&~& 54.308&~&~& $\mathrm{GPa}$ \\ 
		$\mu$ &~&~& 27.174 &~&~& $\mathrm{GPa}$ \\ 
		$l$ &~& ~&-281.5&~&~& $\mathrm{GPa}$ \\ 
		$m$ &~&~& -339&~&~& $\mathrm{GPa}$ \\ 
		$n$ &~&~& -416 &~&~& $\mathrm{GPa}$ \\ \hline
	\end{tabular}
	\label{table:AI_6061}
\end{table}

\subsection{Wave propagation direction parallel to prestress}
In this part, the wave propagation direction is changed to parallel to the prestress. The results of this simulation will be compared with the results of the Effective Elastic Constants (EEC) method used in the past, and the errors of the EEC method will be pointed out. 

The dimensions of the model is the same as in the previous part, the cross-sectional schematic of the plate is shown in Fig. \ref{fig:AI_plate}, but the material parameters are taken from Table \ref{table:AI_120} for comparison with other paper \cite{kubrusly2016time}. The Lamb wave propagates in the $\xi_3$ direction of a plate subjected to a tensile stress of magnitude Siel33 in the $\xi_3$ direction. The prestress tensor can be written as
\begin{equation}
s^i=\begin{pmatrix}0&0&0\\\\0&0&0\\\\0&0&\mathrm{Siel33}\end{pmatrix}
\end{equation} 

\subsubsection{Acoustoelastic SAFE simulation}
\begin{figure}[htbp]
	\centering
	\subfloat[Full frequency range]{\label{subfig:weak_a}\includegraphics[width = 0.5\columnwidth]{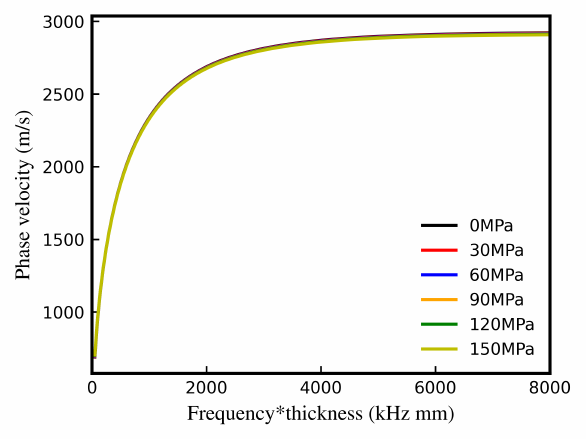}} \\
	\subfloat[Low frequency]{\label{subfig:weak_b}\includegraphics[width = 0.5\columnwidth]{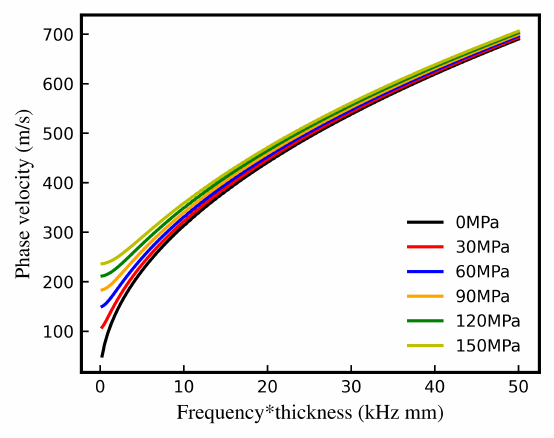}} 
	\subfloat[High frequency]{\label{subfig:weak_c}\includegraphics[width = 0.53\columnwidth]{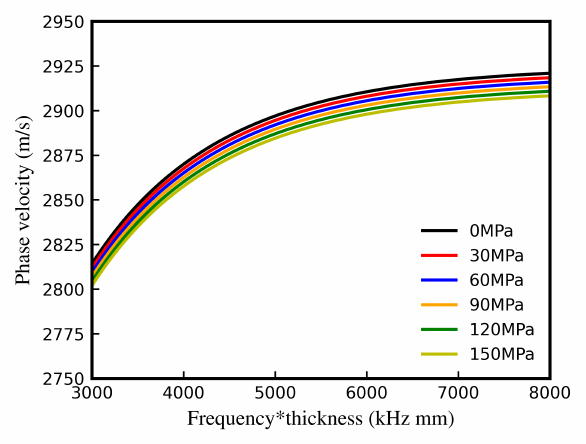}} \\
	\subfloat[Velocity difference at low frequency]{\label{subfig:weak_d}\includegraphics[width = 0.5\columnwidth]{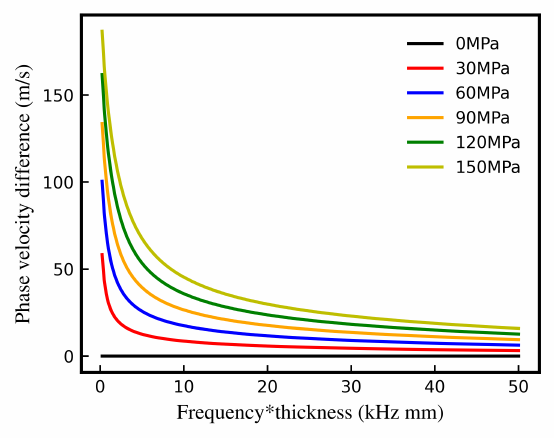}} 
	\subfloat[Velocity difference, zero crossing region]{\label{subfig:weak_e}\includegraphics[width = 0.53\columnwidth]{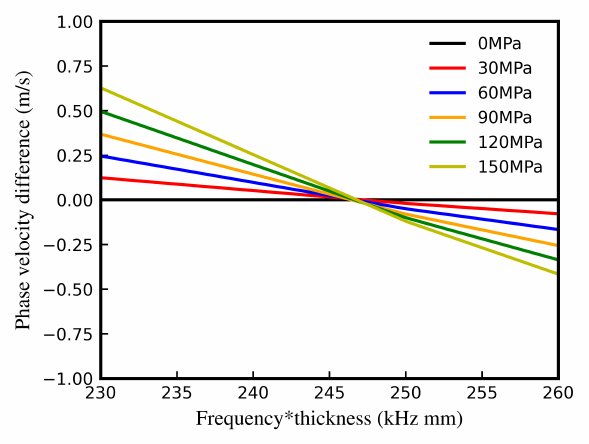}} \\	
	\caption{Phase velocity dispersion curves of A0 Lamb wave under different tensile stresses calculated by the weak form PDE method}
	\label{fig:weak}
\end{figure}

The phase velocity curves of A0 mode are solved using the weak form simulation approach proposed in this paper. The dispersion curves are obtained for prestresses of 0 MPa, 30 MPa, 60 MPa, 90 MPa, 120 MPa, 150 MPa, respectively. The results are shown in Fig. \ref{fig:weak}. Fig. \ref{subfig:weak_a} shows the full frequency range of phase velocity curves under different stresses, while Fig. \ref{subfig:weak_b} and Fig. \ref{subfig:weak_c} show the low frequency and high frequency range respectively. It can be seen from Fig. \ref{subfig:weak_a} that the acoustoelastic effect is very small. It can be seen from Fig. \ref{subfig:weak_b} and Fig. \ref{subfig:weak_c} that at low frequencies, stress has a positive effect on phase velocity, while at high frequencies, it has a negative effect. This means that at low frequencies, phase velocity increases with increasing stress, but at high frequencies, phase velocity decreases as stress increases. This also means that there exists a range in which the stress gain on the phase velocity changes from positive to negative, and this range is called the zero crossing region. Another point of interest is that as the frequency approaches zero, prestress causes the phase velocity to converge to a specific non-zero value, rather than zero, as shown in Fig. \ref{subfig:weak_d}. The difference in phase velocity tends to infinity as the stress increases. The specific zero crossing region is shown in Fig. \ref{subfig:weak_e}. The zero cross frequency-thickness is about  246 kHz $\times$ mm. These conclusions are consistent with the theoretical solutions presented in the paper \cite{kubrusly2016time}.
 
\subsubsection{Effective Elastic Constants method}
To demonstrate the advantages of the weak form PDE method proposed in this paper, the same analysis is performed using the EEC method. The phase velocity dispersion curves of A0 Lamb wave under different tensile stresses calculated by the EEC method. In order to solve the problem of $A^{n}_{\alpha\beta\gamma\delta} \ne A^{n}_{\alpha\beta\delta\gamma}$ and to be able to solve the Lamb wave dispersion curve in commercial software, the EEC method is used \cite{yang2019acoustoelastic}. The EEC method simplifies complexity by splitting the tensor $A^{n}$, two split approaches (EEC1 and EEC2) are proposed in \cite{kubrusly2016time}. The EEC1 case is better in \cite{kubrusly2016time}. When the prestress is 120 MPa, the component selection for EEC1 is given by Eq. (\ref{eq:EEC1}). Note the Voigt notation order in COMSOL.

\begin{figure*}[!ht]
	\centering
	\subfloat[Full frequency range]{\label{subfig:EEC1_a}\includegraphics[width = 0.5\columnwidth]{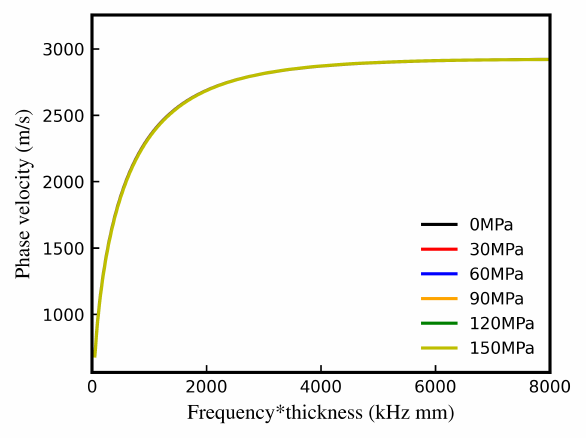}}
	\subfloat[High frequency]{\includegraphics[width = 0.5\columnwidth]{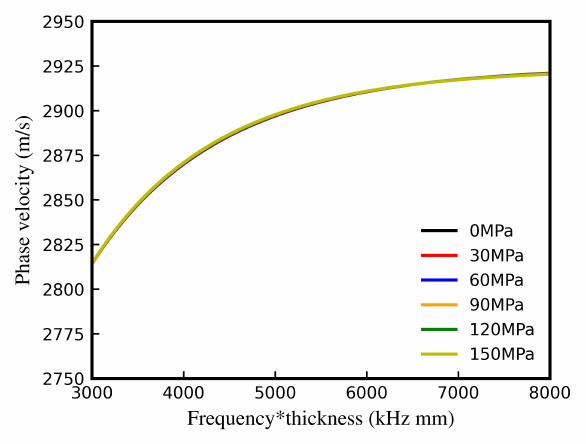}} \\
	\subfloat[Low frequency]{\includegraphics[width = 0.5\columnwidth]{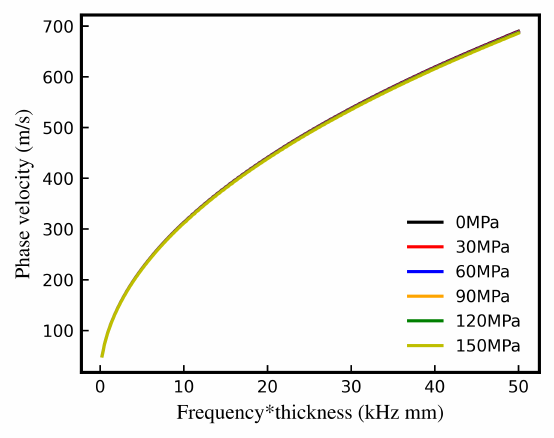}}
	\subfloat[Velocity difference at low frequency]{\label{subfig:EEC1_d}\includegraphics[width = 0.5\columnwidth]{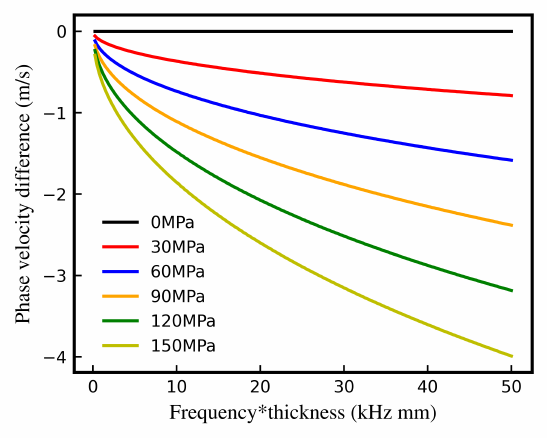}} \\
	\caption{Phase velocity dispersion curves of A0 Lamb wave under different tensile stresses calculated by the EEC1 method}
	\label{fig:EEC1}
\end{figure*}

The results are presented in Fig. \ref{fig:EEC1}. After zooming in the low and high frequency regions, the phase velocity curves for different stresses are still blended. Stress has a negative effect on phase velocity in full frequency range, the phase velocity decreases as the stress increases, which also means that there is no zero crossing region. As the frequency approaches zero, prestress causes the phase velocity to converge to zero, rather than a specific non-zero value, as shown in Fig. \ref{subfig:EEC1_d}.  These results are different from the theoretical solutions in the \cite{kubrusly2016time}. 

By comparing the weak form PDE with the EEC method, it is evident that the method proposed in this paper has a more comprehensive theoretical framework and results that are more in line with the theoretical solutions. Particularly in the low frequency region, the EEC method's solution results are close to errors. 

\begin{equation} \label{eq:EEC1}
	A_{EEC1}^{n} = 
	\bordermatrix{
		&11&22&33&32&31&21 \cr
		11&108.24&55.06&54.51&0&0&0\cr
		22&55.06&108.24&54.51&0&0&0\cr
		33&54.51&54.51&105.91&0&0&0\cr
		32&0&0&0&26.31&0&0\cr
		31&0&0&0&0&26.31&0\cr
		21&0&0&0&0&0&26.59\cr
	} GPa
\end{equation}

\subsection{Acoustoelastic wave propagation simulation}
The theoretical solutions for wave propagation in prestressed media are challenging to obtain. The use of the EEC method to simulate the wave propagation process is discussed in paper \cite{kubrusly2016time}. However, the shortcomings of the EEC method have already been mentioned above. 

\begin{figure}[!ht]
	\centering 
	\includegraphics[width=\columnwidth]{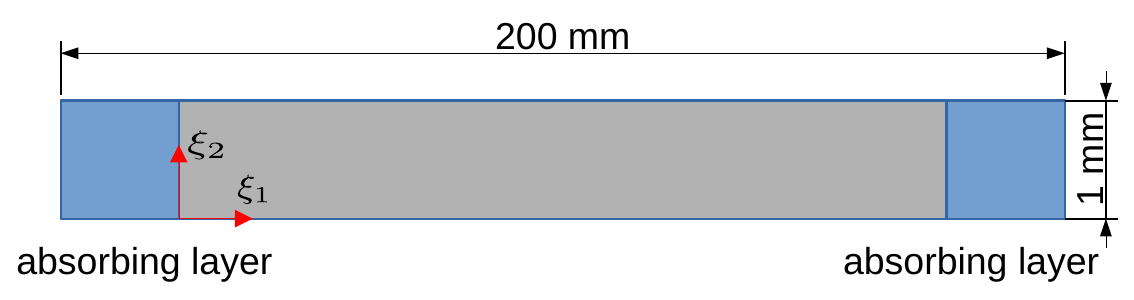} 
	\caption{The schematic of the plate} 
	\label{fig:Lamb_plate}
\end{figure}

\begin{figure}[!ht]
	\centering 
	\includegraphics[width=0.6\columnwidth]{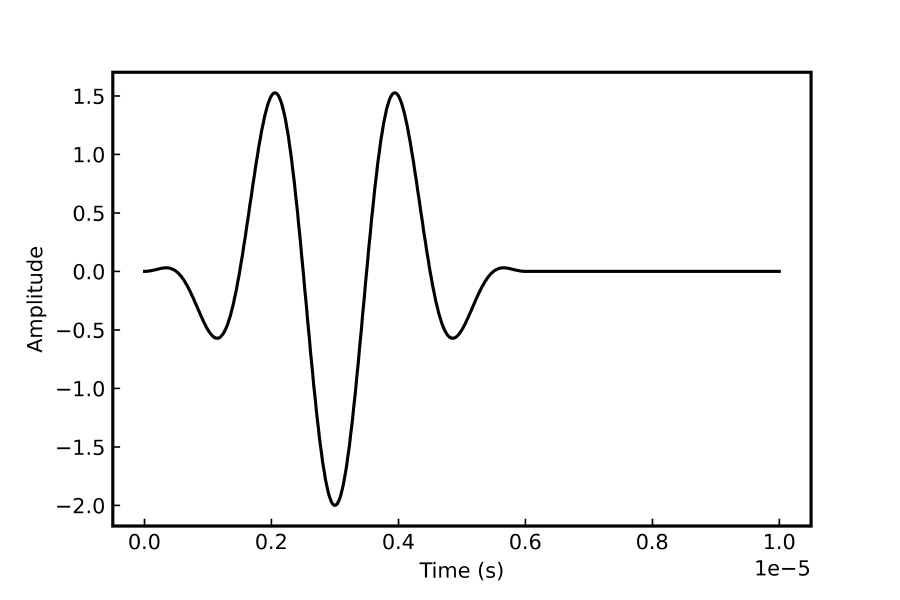} 
	\caption{Time domain curve of excitation wave} 
	\label{fig:incident_wave}
\end{figure}

The A0 mode Lamb wave is excited to simulate the propagation process. The plate is 200 mm long and 1 mm wide, with material parameters as shown in Table \ref{table:AI_120}, there are 20 mm absorbing layers at both ends of the plate. The schematic of the plate is shown in Fig. \ref{fig:Lamb_plate}. At point A (20 mm, 1 mm), a time-dependent out-of-plane displacement \( u_y \) is applied, and the same \( u_y \) is applied at point B (20 mm, 0 mm) to simulate antisymmetric modes. If to simulate symmetric modes, \( u_y \) and \( -u_y \) can be applied at points A and B, respectively. The excitation wave signal is given in the Eq. (\ref{eq:ultrasonic_wave}) and 0.5 MHz is selected as the frequency parameter. The time-dependent out-of-plane exciting wave is shown in Fig. \ref{fig:incident_wave}. The total simulation time is 50 $\mu s$, the time step is 0.025 $\mu s$, and the mesh size is 0.1 mm.

\begin{equation}\label{eq:ultrasonic_wave}
	F(t) = 
	\begin{cases}
		\left[1-\cos(\frac{2\pi f}{3}t)\right] \cos(2\pi ft), \quad &\text{for} \quad 0 \le t \le \frac{3.0}{f}\\
		0.0 &\text{otherwise}
	\end{cases}
\end{equation}

\begin{figure}[!ht]
	\centering 
	\includegraphics[width=0.6\columnwidth]{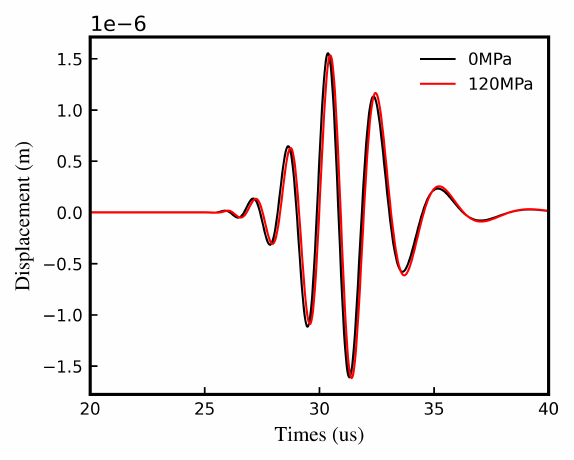} 
	\caption{The influence of prestress on the wave propagation process} 
	\label{fig:dis_time}
\end{figure}

Unstressed and 120 MPa prestressed conditions are applied parallel to the wave propagation direction. The \( u_y \) displacement curves at the coordinate point (100 mm, 1 mm) are extracted, as shown in Fig. \ref{fig:dis_time}. It can also be observed from the Fig. \ref{fig:dis_time} that the influence of stress on the wave amplitude and other characteristics is very small. The wave propagation simulation work in this part provides a reliable foundation for future acoustoelastic applications. 

\section{Conclusion}
\begin{enumerate}
	\itemsep=0pt
	\item A new weak form PDE method is proposed, which can simulate the propagation process of waves in prestressed media or the dispersion curve of Lamb wave;
	\item The new method is used to simulate the dispersion curves in two cases: when the propagation direction is perpendicular to the prestress and when it is parallel to the prestress. In the case when the propagation direction is perpendicular to the prestress, the phase velocity difference curve caused by prestress is compared with the results from the SPBW method. The agreement between the two confirms the correctness of the theory and simulations presented in this paper. In the case when the propagation direction is parallel to the prestress, the variation curve of phase velocity with stress is compared with the EEC method. Several issues with the EEC method are highlighted, such as the convergence value at zero frequency, the positive and negative gains of stress at low and high frequencies, and the absence of zero crossing region. The method proposed in this paper perfectly aligns with the theoretical solution;
	\item The propagation of A0 mode Lamb wave in a plate is simulated, demonstrating the small effect of prestress on the wave. This provides a basis for subsequent studies;
\end{enumerate}  

\appendix
\section{Appendix}
The components of $P^n_{\alpha\beta}$ and $M^n_{\alpha}$ are shown here. Most variables, such as \( ux, l \), are the same as mentioned in the main text. nx, ny and nz represent the normal vector X-component, Y-component, Z-component, respectively. 
\begin{equation}
\begin{aligned}
\mathrm{P11} &= ux(u^i_{1,1}(2l + 4.0m) + 3u^i_{1,1}(\lambda + 2\mu) + 2u^i_{2,2}l + u^i_{2,2}\lambda + 2u^i_{3,3}l + u^i_{3,3}\lambda + \lambda + 2\mu) + uy(u^i_{1,2}\lambda + 2u^i_{1,2}\mu + u^i_{2,1}\mu + 2.0m\\&(u^i_{1,2}/2 + u^i_{2,1}/2)) + uz(u^i_{1,3}\lambda + 2u^i_{1,3}\mu + u^i_{3,1}\mu + 2.0m(u^i_{1,3}/2 + u^i_{3,1}/2)) + vx(u^i_{1,2}\mu + u^i_{2,1}(\lambda + 2\mu) \\&+ 2.0m(u^i_{1,2}/2 + u^i_{2,1}/2)) + vy(2u^i_{1,1}l + u^i_{1,1}\lambda + 2u^i_{2,2}l + u^i_{2,2}\lambda + u^i_{3,3}(2l - 2m + n) + \lambda) + vz(u^i_{2,3}\lambda + 2(u^i_{2,3}/2 \\&+ u^i_{3,2}/2)(1.0m - 0.5n)) + wx(u^i_{1,3}\mu + u^i_{3,1}(\lambda + 2\mu) + 2.0m(u^i_{1,3}/2 + u^i_{3,1}/2)) + wy(u^i_{3,2}\lambda + 2(u^i_{2,3}/2 + u^i_{3,2}/2)\\&(1.0m - 0.5n)) + wz(2u^i_{1,1}l + u^i_{1,1}\lambda + u^i_{2,2}(2l - 2m + n) + 2u^i_{3,3}l + u^i_{3,3}\lambda + \lambda) \\
\mathrm{P12} &= ux(u^i_{1,2}\lambda + 2u^i_{1,2}\mu + u^i_{2,1}\mu + 2.0m(u^i_{1,2}/2 + u^i_{2,1}/2)) + uy(1.0u^i_{1,1}m + u^i_{1,1}\lambda + 2u^i_{1,1}\mu + 1.0u^i_{2,2}m + u^i_{2,2}(\lambda + 2\mu) + \\&u^i_{3,3}\lambda + u^i_{3,3}(1.0m - 0.5n) + \mu) + uz(u^i_{2,3}\mu + u^i_{3,2}\mu + 0.5n(u^i_{2,3}/2 + u^i_{3,2}/2)) + vx(1.0u^i_{1,1}m + u^i_{1,1}\mu + 1.0u^i_{2,2}m + u^i_{2,2}\mu \\&+ u^i_{3,3}(1.0m - 0.5n) + \mu) + vy(u^i_{1,2}(\lambda + 2\mu) + u^i_{2,1}\mu + 2.0m(u^i_{1,2}/2 + u^i_{2,1}/2)) + vz(u^i_{1,3}\mu + 0.5n(u^i_{1,3}/2 + u^i_{3,1}/2)) \\&+ wx(u^i_{3,2}\mu + 0.5n(u^i_{2,3}/2 + u^i_{3,2}/2)) + wy(u^i_{1,3}\mu + u^i_{3,1}\mu + 0.5n(u^i_{1,3}/2 + u^i_{3,1}/2)) + wz(u^i_{1,2}\lambda + 2(u^i_{1,2}/2 \\&+ u^i_{2,1}/2)(1.0m - 0.5n)) \\
\mathrm{P13} &= ux(u^i_{1,3}\lambda + 2u^i_{1,3}\mu + u^i_{3,1}\mu + 2.0m(u^i_{1,3}/2 + u^i_{3,1}/2)) + uy(u^i_{2,3}\mu + u^i_{3,2}\mu + 0.5n(u^i_{2,3}/2 + u^i_{3,2}/2)) + uz(1.0u^i_{1,1}m \\&+ u^i_{1,1}\lambda + 2u^i_{1,1}\mu + u^i_{2,2}\lambda + u^i_{2,2}(1.0m - 0.5n) + 1.0u^i_{3,3}m + u^i_{3,3}(\lambda + 2\mu) + \mu) + vx(u^i_{2,3}\mu + 0.5n(u^i_{2,3}/2 + u^i_{3,2}/2)) \\&+ vy(u^i_{1,3}\lambda + 2(u^i_{1,3}/2 + u^i_{3,1}/2)(1.0m - 0.5n)) + vz(u^i_{1,2}\mu + u^i_{2,1}\mu + 0.5n(u^i_{1,2}/2 + u^i_{2,1}/2)) + wx(1.0u^i_{1,1}m + u^i_{1,1}\mu \\&+ u^i_{2,2}(1.0m - 0.5n) + 1.0u^i_{3,3}m + u^i_{3,3}\mu + \mu) + wy(u^i_{1,2}\mu + 0.5n(u^i_{1,2}/2 + u^i_{2,1}/2)) + wz(u^i_{1,3}(\lambda + 2\mu) + u^i_{3,1}\mu \\&+ 2.0m(u^i_{1,3}/2 + u^i_{3,1}/2)) \\
\mathrm{P21} &= ux(u^i_{1,2}\mu + u^i_{2,1}(\lambda + 2\mu) + 2.0m(u^i_{1,2}/2 + u^i_{2,1}/2)) + uy(1.0u^i_{1,1}m + u^i_{1,1}\mu + 1.0u^i_{2,2}m + u^i_{2,2}\mu + u^i_{3,3}(1.0m - 0.5n) + \mu) \\&+ uz(u^i_{2,3}\mu + 0.5n(u^i_{2,3}/2 + u^i_{3,2}/2)) + vx(1.0u^i_{1,1}m + u^i_{1,1}(\lambda + 2\mu) + 1.0u^i_{2,2}m + u^i_{2,2}\lambda + 2u^i_{2,2}\mu + u^i_{3,3}\lambda + u^i_{3,3}(1.0m \\&- 0.5n) + \mu) + vy(u^i_{1,2}\mu + u^i_{2,1}\lambda + 2u^i_{2,1}\mu + 2.0m(u^i_{1,2}/2 + u^i_{2,1}/2)) + vz(u^i_{1,3}\mu + u^i_{3,1}\mu + 0.5n(u^i_{1,3}/2 + u^i_{3,1}/2)) \\&+ wx(u^i_{2,3}\mu + u^i_{3,2}\mu + 0.5n(u^i_{2,3}/2 + u^i_{3,2}/2)) + wy(u^i_{3,1}\mu + 0.5n(u^i_{1,3}/2 + u^i_{3,1}/2)) + wz(u^i_{2,1}\lambda + 2(u^i_{1,2}/2 \\&+ u^i_{2,1}/2)(1.0m - 0.5n)) \\
\mathrm{P22} & = ux(2u^i_{1,1}l + u^i_{1,1}\lambda + 2u^i_{2,2}l + u^i_{2,2}\lambda + u^i_{3,3}(2l - 2m + n) + \lambda) + uy(u^i_{1,2}(\lambda + 2\mu) + u^i_{2,1}\mu + 2.0m(u^i_{1,2}/2 + u^i_{2,1}/2)) \\&+ uz(u^i_{1,3}\lambda + 2(u^i_{1,3}/2 + u^i_{3,1}/2)(1.0m - 0.5n)) + vx(u^i_{1,2}\mu + u^i_{2,1}\lambda + 2u^i_{2,1}\mu + 2.0m(u^i_{1,2}/2 + u^i_{2,1}/2)) + vy(2u^i_{1,1}l \\&+ u^i_{1,1}\lambda + u^i_{2,2}(2l + 4.0m) + 3u^i_{2,2}(\lambda + 2\mu) + 2u^i_{3,3}l + u^i_{3,3}\lambda + \lambda + 2\mu) + vz(u^i_{2,3}\lambda + 2u^i_{2,3}\mu + u^i_{3,2}\mu + 2.0m(u^i_{2,3}/2 \\&+ u^i_{3,2}/2)) + wx(u^i_{3,1}\lambda + 2(u^i_{1,3}/2 + u^i_{3,1}/2)(1.0m - 0.5n)) + wy(u^i_{2,3}\mu + u^i_{3,2}(\lambda + 2\mu) + 2.0m(u^i_{2,3}/2 + u^i_{3,2}/2)) \\&+ wz(u^i_{1,1}(2l - 2m + n) + 2u^i_{2,2}l + u^i_{2,2}\lambda + 2u^i_{3,3}l + u^i_{3,3}\lambda + \lambda) \\
\mathrm{P23} & = ux(u^i_{2,3}\lambda + 2(u^i_{2,3}/2 + u^i_{3,2}/2)(1.0m - 0.5n)) + uy(u^i_{1,3}\mu + 0.5n(u^i_{1,3}/2 + u^i_{3,1}/2)) + uz(u^i_{1,2}\mu + u^i_{2,1}\mu + 0.5n(u^i_{1,2}/2 \\&+ u^i_{2,1}/2)) + vx(u^i_{1,3}\mu + u^i_{3,1}\mu + 0.5n(u^i_{1,3}/2 + u^i_{3,1}/2)) + vy(u^i_{2,3}\lambda + 2u^i_{2,3}\mu + u^i_{3,2}\mu + 2.0m(u^i_{2,3}/2 + u^i_{3,2}/2)) \\&+ vz(u^i_{1,1}\lambda + u^i_{1,1}(1.0m - 0.5n) + 1.0u^i_{2,2}m + u^i_{2,2}\lambda + 2u^i_{2,2}\mu + 1.0u^i_{3,3}m + u^i_{3,3}(\lambda + 2\mu) + \mu) + wx(u^i_{2,1}\mu + 0.5n(u^i_{1,2}/2 \\&+ u^i_{2,1}/2)) + wy(u^i_{1,1}(1.0m - 0.5n) + 1.0u^i_{2,2}m + u^i_{2,2}\mu + 1.0u^i_{3,3}m + u^i_{3,3}\mu + \mu) + wz(u^i_{2,3}(\lambda + 2\mu) + u^i_{3,2}\mu + 2.0m(u^i_{2,3}/2 \\&+ u^i_{3,2}/2)) \\
 \nonumber
\end{aligned}
\end{equation}
\begin{equation}
\begin{aligned}
\mathrm{P31} & = ux(u^i_{1,3}\mu + u^i_{3,1}(\lambda + 2\mu) + 2.0m(u^i_{1,3}/2 + u^i_{3,1}/2)) + uy(u^i_{3,2}\mu + 0.5n(u^i_{2,3}/2 + u^i_{3,2}/2)) + uz(1.0u^i_{1,1}m + u^i_{1,1}\mu \\&+ u^i_{2,2}(1.0m - 0.5n) + 1.0u^i_{3,3}m + u^i_{3,3}\mu + \mu) + vx(u^i_{2,3}\mu + u^i_{3,2}\mu + 0.5n(u^i_{2,3}/2 + u^i_{3,2}/2)) + vy(u^i_{3,1}\lambda + 2(u^i_{1,3}/2 \\&+ u^i_{3,1}/2)(1.0m - 0.5n)) + vz(u^i_{2,1}\mu + 0.5n(u^i_{1,2}/2 + u^i_{2,1}/2)) + wx(1.0u^i_{1,1}m + u^i_{1,1}(\lambda + 2\mu) + u^i_{2,2}\lambda + u^i_{2,2}(1.0m \\&- 0.5n) + 1.0u^i_{3,3}m + u^i_{3,3}\lambda + 2u^i_{3,3}\mu + \mu) + wy(u^i_{1,2}\mu + u^i_{2,1}\mu + 0.5n(u^i_{1,2}/2 + u^i_{2,1}/2)) + wz(u^i_{1,3}\mu + u^i_{3,1}\lambda + 2u^i_{3,1}\mu \\&+ 2.0m(u^i_{1,3}/2 + u^i_{3,1}/2)) \\
\mathrm{P32} & = ux(u^i_{3,2}\lambda + 2(u^i_{2,3}/2 + u^i_{3,2}/2)(1.0m - 0.5n)) + uy(u^i_{1,3}\mu + u^i_{3,1}\mu + 0.5n(u^i_{1,3}/2 + u^i_{3,1}/2)) + uz(u^i_{1,2}\mu + 0.5n(u^i_{1,2}/2 \\&+ u^i_{2,1}/2)) + vx(u^i_{3,1}\mu + 0.5n(u^i_{1,3}/2 + u^i_{3,1}/2)) + vy(u^i_{2,3}\mu + u^i_{3,2}(\lambda + 2\mu) + 2.0m(u^i_{2,3}/2 + u^i_{3,2}/2)) + vz(u^i_{1,1}(1.0m \\&- 0.5n) + 1.0u^i_{2,2}m + u^i_{2,2}\mu + 1.0u^i_{3,3}m + u^i_{3,3}\mu + \mu) + wx(u^i_{1,2}\mu + u^i_{2,1}\mu + 0.5n(u^i_{1,2}/2 + u^i_{2,1}/2)) + wy(u^i_{1,1}\lambda \\&+ u^i_{1,1}(1.0m - 0.5n) + 1.0u^i_{2,2}m + u^i_{2,2}(\lambda + 2\mu) + 1.0u^i_{3,3}m + u^i_{3,3}\lambda + 2u^i_{3,3}\mu + \mu) + wz(u^i_{2,3}\mu + u^i_{3,2}\lambda + 2u^i_{3,2}\mu \\&+ 2.0m(u^i_{2,3}/2 + u^i_{3,2}/2)) \\
\mathrm{P33} & = ux(2u^i_{1,1}l + u^i_{1,1}\lambda + u^i_{2,2}(2l - 2m + n) + 2u^i_{3,3}l + u^i_{3,3}\lambda + \lambda) + uy(u^i_{1,2}\lambda + 2(u^i_{1,2}/2 + u^i_{2,1}/2)(1.0m - 0.5n)) \\&+ uz(u^i_{1,3}(\lambda + 2\mu) + u^i_{3,1}\mu + 2.0m(u^i_{1,3}/2 + u^i_{3,1}/2)) + vx(u^i_{2,1}\lambda + 2(u^i_{1,2}/2 + u^i_{2,1}/2)(1.0m - 0.5n)) + vy(u^i_{1,1}(2l \\&- 2m + n) + 2u^i_{2,2}l + u^i_{2,2}\lambda + 2u^i_{3,3}l + u^i_{3,3}\lambda + \lambda) + vz(u^i_{2,3}(\lambda + 2\mu) + u^i_{3,2}\mu + 2.0m(u^i_{2,3}/2 + u^i_{3,2}/2)) + wx(u^i_{1,3}\mu \\&+ u^i_{3,1}\lambda + 2u^i_{3,1}\mu + 2.0m(u^i_{1,3}/2 + u^i_{3,1}/2)) + wy(u^i_{2,3}\mu + u^i_{3,2}\lambda + 2u^i_{3,2}\mu + 2.0m(u^i_{2,3}/2 + u^i_{3,2}/2)) + wz(2u^i_{1,1}l \\&+ u^i_{1,1}\lambda + 2u^i_{2,2}l + u^i_{2,2}\lambda + u^i_{3,3}(2l + 4.0m) + 3u^i_{3,3}(\lambda + 2\mu) + \lambda + 2\mu) \\
\mathrm{M1} & = nx(u^i_{1,1}\lambda vy + u^i_{1,1}\lambda wz + u^i_{1,2}\mu vx + u^i_{1,3}\mu wx + ux(2u^i_{1,1}(\lambda + 2\mu) + u^i_{2,2}\lambda + u^i_{3,3}\lambda) + uy(2u^i_{1,2}\mu + u^i_{2,1}\mu) + uz(2u^i_{1,3}\mu \\&+ u^i_{3,1}\mu)) + ny(u^i_{1,1}\mu vx + u^i_{1,2}\lambda wz + u^i_{1,2}vy(\lambda + 2\mu) + u^i_{1,3}\mu vz + u^i_{1,3}\mu wy + ux(u^i_{1,2}\lambda + u^i_{1,2}\mu + u^i_{2,1}\mu) + uy(u^i_{1,1}\lambda \\&+ u^i_{1,1}\mu + u^i_{2,2}(\lambda + 2\mu) + u^i_{3,3}\lambda) + uz(u^i_{2,3}\mu + u^i_{3,2}\mu)) + nz(u^i_{1,1}\mu wx + u^i_{1,2}\mu vz + u^i_{1,2}\mu wy + u^i_{1,3}\lambda vy + u^i_{1,3}wz(\lambda + 2\mu) \\&+ ux(u^i_{1,3}\lambda + u^i_{1,3}\mu + u^i_{3,1}\mu) + uy(u^i_{2,3}\mu + u^i_{3,2}\mu) + uz(u^i_{1,1}\lambda + u^i_{1,1}\mu + u^i_{2,2}\lambda + u^i_{3,3}(\lambda + 2\mu))) \\
\mathrm{M2} & = nx(u^i_{2,1}\lambda wz + u^i_{2,1}ux(\lambda + 2\mu) + u^i_{2,2}\mu uy + u^i_{2,3}\mu uz + u^i_{2,3}\mu wx + vx(u^i_{1,1}(\lambda + 2\mu) + u^i_{2,2}\lambda + u^i_{2,2}\mu + u^i_{3,3}\lambda) + vy(u^i_{1,2}\mu \\&+ u^i_{2,1}\lambda + u^i_{2,1}\mu) + vz(u^i_{1,3}\mu + u^i_{3,1}\mu)) + ny(u^i_{2,1}\mu uy + u^i_{2,2}\lambda ux + u^i_{2,2}\lambda wz + u^i_{2,3}\mu wy + vx(u^i_{1,2}\mu + 2u^i_{2,1}\mu) + vy(u^i_{1,1}\lambda \\&+ 2u^i_{2,2}(\lambda + 2\mu) + u^i_{3,3}\lambda) + vz(2u^i_{2,3}\mu + u^i_{3,2}\mu)) + nz(u^i_{2,1}\mu uz + u^i_{2,1}\mu wx + u^i_{2,2}\mu wy + u^i_{2,3}\lambda ux + u^i_{2,3}wz(\lambda + 2\mu) \\&+ vx(u^i_{1,3}\mu + u^i_{3,1}\mu) + vy(u^i_{2,3}\lambda + u^i_{2,3}\mu + u^i_{3,2}\mu) + vz(u^i_{1,1}\lambda + u^i_{2,2}\lambda + u^i_{2,2}\mu + u^i_{3,3}(\lambda + 2\mu))) \\
\mathrm{M3} & = nx(u^i_{3,1}\lambda vy + u^i_{3,1}ux(\lambda + 2\mu) + u^i_{3,2}\mu uy + u^i_{3,2}\mu vx + u^i_{3,3}\mu uz + wx(u^i_{1,1}(\lambda + 2\mu) + u^i_{2,2}\lambda + u^i_{3,3}\lambda + u^i_{3,3}\mu) + wy(u^i_{1,2}\mu \\&+ u^i_{2,1}\mu) + wz(u^i_{1,3}\mu + u^i_{3,1}\lambda + u^i_{3,1}\mu)) + ny(u^i_{3,1}\mu uy + u^i_{3,1}\mu vx + u^i_{3,2}\lambda ux + u^i_{3,2}vy(\lambda + 2\mu) + u^i_{3,3}\mu vz + wx(u^i_{1,2}\mu \\&+ u^i_{2,1}\mu) + wy(u^i_{1,1}\lambda + u^i_{2,2}(\lambda + 2\mu) + u^i_{3,3}\lambda + u^i_{3,3}\mu) + wz(u^i_{2,3}\mu + u^i_{3,2}\lambda + u^i_{3,2}\mu)) + nz(u^i_{3,1}\mu uz + u^i_{3,2}\mu vz \\&+ u^i_{3,3}\lambda ux + u^i_{3,3}\lambda vy + wx(u^i_{1,3}\mu + 2u^i_{3,1}\mu) + wy(u^i_{2,3}\mu + 2u^i_{3,2}\mu) + wz(u^i_{1,1}\lambda + u^i_{2,2}\lambda + 2u^i_{3,3}(\lambda + 2\mu)))
 \nonumber
\end{aligned}
\end{equation}

\printcredits

\section*{Declaration of competing interest}
The authors declare that they have no known competing financial interests or personal relationships that could have appeared to influence the work reported in this paper.
\section*{Data availability}
Data will be made available on request.
\section*{Acknowledgements}
This study is supported by the National Natural Science Foundation of China Project (Grant No. 51771051), the Natural Science Foundation of Liaoning Province Project (Grant No. 2021-MS-102) and the Fundamental Research Funds for the Central Universities (Grant No. N2105021).
\bibliographystyle{model1-num-names}
\bibliography{cas-refs}
\end{document}